\documentclass[11pt,twoside]{article}  
\usepackage{apn3conf}

\begin{document}   

%
%
%
%

\title{The Egg Nebula (AFGL 2688): Deepening Enigma }
\titlemark{ The Egg Nebula (AFGL 2688)}

%
%
%

\author{Joel H. Kastner }
\affil{Center for Imaging Science, Rochester Institute of
  Technology, Rochester, NY 14623, USA; jhk@cis.rit.edu}
\author{Noam Soker }
\affil{Department of Physics, Technion$-$Israel institute of Technology,
Haifa 32000, Israel, and Department of Physics, Oranim, Israel;
  soker@physics.technion.ac.il}

%
%

\contact{Joel Kastner }
\email{jhk@cis.rit.edu }

%
%
%
%
%

\paindex{Kastner, J.H. }
\aindex{Soker, N. }     

%
%

\authormark{ Kastner \& Soker}

%
%



\begin{abstract}          

Recent observations of the Egg Nebula (AFGL 2688), obtained
at ever-increasing spatial and spectral resolution, have
revealed a perplexing array of phenomena. Many of these
phenomena present challenges to our understanding of this
object as an emerging, bipolar planetary nebula. Here, we
consider two particularly intriguing aspects of the Egg: the
peculiar structure and kinematics of its equatorial regions,
and the nature of an apparent widely separated companion to
the central star. In the first case, we
use recently acquired {\it Hubble Space Telescope} images to
demonstrate that the 
H$_2$ emission distributed east and west of the central star
is spatially coincident with a dusty, equatorial disk or
torus. The H$_2$ is thus constrained to lie near the equatorial plane,
casting doubt on pure radial outflow models for the
equatorial kinematics. In the second case, we show that the
apparent companion (``Peak A'') may be an accreting white
dwarf that has undergone one or more thermonuclear bursts.

\end{abstract}

%
%

\section{Key Problems Posed by the Egg}

The Egg Nebula (AFGL 2688) has long been regarded as
exemplary of objects in transition from asymptotic giant branch (AGB)
star to bipolar planetary nebula (PN). 
Recent optical, infrared, and radio observations at high
spatial and spectral resolution have revealed a remarkable
degree of complexity in this object (for a brief
synopsis, see Kastner et al.\ 2002). We examine here
two particularly puzzling and controversial aspects of AFGL
2688 that may provide clues to the origin of axial symmetry
in PNs and proto-PNs. \\
1. An equatorial velocity gradient of magnitude
  $\sim10-20$ km s$^{-1}$ --- similar to that observed along
  the polar axis --- has been detected in various molecular
  tracers. This gradient has been attributed both
  to multiple, radial outflows (Cox et al.\ 2000) and to a
  component of azimuthal velocity (i.e., rotation about the
  polar axis; Kastner et al.\ 2001). Each
  model has problems. The ``pure radial outflow'' model
  does not appear viable if the emitting molecular gas is
  indeed confined to the equatorial plane, while the ``rotation'' model
  requires an extremely (untenably?) large reservoir of angular
  momentum. \\
2. Near-infrared polarimetric imaging suggests that a
  luminous source of direct emission lies
  embedded very near ($\sim0.5''$ from) the central,
  illuminating star. This object therefore likely
  constitutes a widely
  separated ($a\approx500$ AU) companion (Sahai et al. 1998;
  Weintraub et al. 2000), although Goto
  et al.\ (2002) argue, on the basis of infrared spectroscopy, that
  the source is instead a knot of reflecting dust.

\section{The Equatorial Region: New Revelations from HST}

\begin{figure}
\plotone{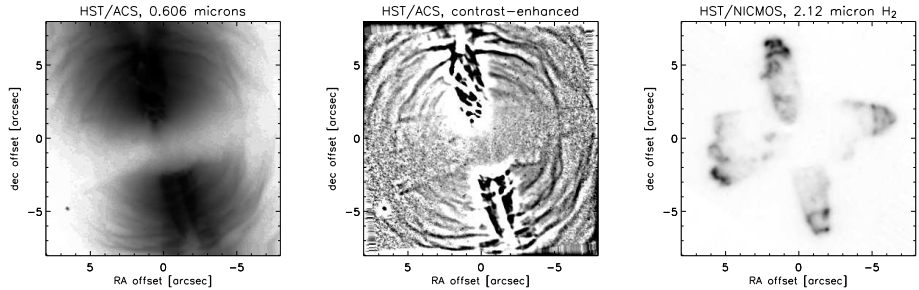}
\caption{Left: Central region of HST/ACS 0.606 $\mu$m image of AFGL
  2688. RA, dec offsets are with respect to the approximate
  position of the central source of illumination. Center:
  the same image, processed via an unsharp  
  masking technique (following Sahai \& Trauger 1998). Right: HST/NICMOS
  2.12 $\mu$m image of H$_2$ emission from AFGL 2688 (Sahai et
  al.\ 1998).}
\end{figure}

Recently acquired {\it Hubble Space Telescope} (HST) images
of AFGL 2688, obtained during 
tests of the new Advanced Camera for Surveys (ACS) in 2002
October, appear to shed new light on the structure of the
equatorial region of the nebula. The images were obtained
with the ACS Wide Field Camera (WFC) through polarimetric
filters and a broad-band filter centered at 0.606 $\mu$m;
the pixel scale of the WFC is $0.05''$ pixel$^{-1}$. Here, we have
averaged the polarized images to form a single, total
intensity image of AFGL 2688.

The left panel of Fig.\ 1 displays the central region of
AFGL 2688 in the ACS/WFC image. These images reveal, in
unprecedented detail, fine structure in the ``dark lane''
that marks the equatorial plane of the system. Of particular
note is the sharp, sculpted southern boundary of the dark
lane. This appearance suggests that the inner region of the
southern reflection lobe is being occulted by material that
is largely confined to the equatorial plane. {\it Provided
that the equatorial regions harbor a dusty disk or torus
that is several arcsec (thousands of AU) in radius,} such
occultation is a natural consequence of the fact that the
north lobe is pointed slightly toward, and the south lobe
slightly away from, the observer (Sahai et al.\ 1998,
Kastner et al.\ 2001, and references therein).

The processed image in the center panel of Fig.\ 1
emphasizes the fine structure in the equatorial regions and,
in particular, reveals the
faint outline of the equatorial torus. The outline of the
torus is most apparent 
along the sharp southern edge of the dark lane,
but also appears as a linear feature extending ``behind''
the inner north lobe. The latter feature could be
interpreted as the outline of the rear-facing edge of the
equatorial disk of AFGL 2688.

There is a dramatic correspondence between
the faint outline of the equatorial torus in the ACS/WFC
image and the distribution of near-IR H$_2$ emission (Fig.\
1, right). In particular, the southern edge of the ``equatorial''
H$_2$ emission (that is, the H$_2$ emission projected along
a line nearly perpendicular to the polar lobes) follows the
undulations in the southern edge 
of the dark lane in exquisite detail. Individual H$_2$ knots
to the southeast and southwest of the position of the central star
have distinct counterparts in the ACS/WFC image. We conclude
that the H$_2$ and the obscuring dust are spatially
coincident; that is, {\it the H$_2$ emission traces the
equatorial plane of the AFGL 2688 system}.

As the near-infrared H$_2$ emission east and west of the
central star is, evidently, confined to the equatorial
regions of AFGL 2688, we conclude that a kinematical model
invoking pure radial outflow cannot explain the large-scale
east-west velocity gradient previously measured in this and
other molecular tracers (wherein emission to the east of the
star is predominantly blueshifted, and emission to the west
predominantly redshifted; Cox et al.\ 2000, Kastner et al.\
2001, and references therein). Alternative models must be
considered. It is conceivable, for example, that the
equatorial H$_2$ and CO emission traces molecular jets, or
``bullets,'' that are initially directed radially outward
along the equator of the system and then are redirected
toward and away from the observer on the east and west sides
of the central star, respectively, by density gradients
within the dusty torus. We note, however, that new H$_2$ and
CO velocity mapping observations of AFGL 2688 at superior spatial
resolution (P. Cox, these proceedings) must be carefully
analyzed to determine the detailed structure of the
equatorial kinematics.

\section{Peak A: an Accreting White Dwarf?}

We propose that the apparent companion
to the central star of AFGL 2688 (``Peak A''; Weintraub et
al.\ 2000) --- if it is indeed a self-luminous object ---
may be an accreting white dwarf (WD) which has undergone one
or more thermonuclear bursts. This companion cannot have 
influenced the formation of bipolar structure in the Egg, as
this process requires a close ($a \stackrel{<}{\sim} 50$ AU)
binary system. Thus Peak A, at a projected orbital
separation of $a\approx 500$ AU, most
likely would constitute a tertiary member of the progenitor system.

To evaluate the possibility that Peak A is an accreting WD,
we first demonstrate that the mass  
accretion rate by a WD companion at such an orbital 
separation can provide the requisite
mass for a thermonuclear burst(s). 
At such large separations the relative velocity between the
accreting companion and the wind, $v_r$, is very nearly the 
speed of the slow wind, $v_s$, where $v_s \approx 10$ km s$^{-1}$
along the equatorial plane of AFGL 2688 (Kastner et al.\ 2001).
 It is reasonable to scale the WD mass assuming
$M_{\rm WD} \stackrel{>}{\sim} 0.6 M_{\odot}$,
as lower mass WDs result either from lower
mass stars --- which is unlikely for the expired companion to
the central star of the Egg Nebula --- or from binary interaction. 
The Bondi-Hoyle accretion rate is then  
$$ 
\dot M \simeq
 10^{-8}
\left( \frac {M_{\rm WD}}{0.6\, M_\odot} \right)^{2}
\left( \frac {v_s}{10\, {\rm km}\, {\rm s}^{-1}} \right)^{-4}
\left( \frac {a}{500\, {\rm AU}} \right)^{-2}
\left( \frac {\vert \dot M_1 \vert }{10^{-4}\, M_\odot\,
 {\rm yr}^{-1}} \right) 
M_\odot \, {\rm yr}^{-1}.
$$ 
The present mass loss rate of the central, illuminating (F
supergiant) star is expected to 
be well below $\dot M_1= 10^{-4} \, M_\odot$ \, yr$^{-1}$.
However, in our scenario the accreting WD went through a 
burst in the last $<300$ yr. 
Because there is a time delay between a mass loss epoch and 
the accretion epoch of that mass, which for the scaling used above is
$t_d = a/v_s = 240$ yr, and the
F supergiant left the AGB $\sim 200$ years ago
(Jura \& Kroto 1990), the WD could still be accreting at present. 

A WD of mass $M_{\rm WD} = 0.6 M_\odot$ that accretes at a rate of
$\sim 10^{-9}$ to $10^{-8} \, M_\odot\,~{\rm yr}^{-1}$ will undergo a 
thermonuclear burst after accreting a mass of $\sim 10^{-4}$
$M_\odot$
(e.g., Fujimoto 1982; Iben 1982; Prialnik \& Kovetz 1995).
From the equation above we find that the Peak A star could have
accreted $\sim 10^{-4}$ of the mass lost by the progenitor
of the Egg Nebula. 
 For an envelope mass of $\sim 1$~$M_\odot$ on the upper AGB,
when the wind speed is low enough to enable substantial
accretion, we find the total accreted mass to be $\sim
10^{-4}$~$M_\odot$, as required
for the burst to occur.
We conclude that the Peak A star, if a WD, has plausibly gone through
at least one burst. Further details concerning this model
will be
presented in Kastner \& Soker (2003, in preparation). 

%
%
%
%

\acknowledgements{We thank Howard Bond and Zolt Levay of
Space Telescope Science Institute for providing FITS
versions of the new HST/ACS images of the Egg.}


\end{document}